\documentstyle{article}

\begin{document}

\def\levelset{{\cal X}}
\def\fxx{{\bf x}}
\def\fxy{{\bf y}}
\def\real{I\!\!R}

\begin{center}
{\LARGE Restriction Conditions of Motion Equations of Non-Abelian }\\
{\LARGE Gauge Fields and Their Wave Solutions }\\
\vskip 0.2in
{\large Mei Xiaochun}\\
\vskip 0.2in
\par
(Department of Physics, Fuzhou University, Fuzhou, 350025, China, E-mail: fzbgk@pub3.fz.fj.cn)
\end{center}
\begin{abstract}
\end{abstract}
\par
Abstract Similar to introduce the Lorentz condition in the motion equation of electromagnetic field, the restriction condition is introduced in the motion equations of non-Ablian gauge fields so that the equations are simplified greatly and their wave solutions are obtained. Only in this way, the Feynman rule's use become rightful in the calculation of perturbation theory.
\par
{\bf PACS numbers:}
1110, 1130 
$$
$$
\par
Let's first review the motion equation of electromagnetic field and its solution. The Lagrangian of free electromagnetic field is
\begin{equation}
L=-{1\over{4}}{F}_{\mu\nu}(x){F}_{\mu\nu}(x)
\end{equation}
Here
\begin{equation}
F_{\mu\nu}(x)=\partial_{\mu}{A}_{\nu}(x)-\partial_{\nu}{A}_{\mu}(x)
\end{equation}
From Eq.(1), the motion equation of free electromagnetic field can be obtained
\begin{equation}
\partial_{\mu}{F}_{\mu\nu}=\partial^2{A}_{\nu}-\partial_{\nu}\partial_{\mu}{A}_{\mu}=0
\end{equation}
On the other hand, Eq.(3) is not a wave equation, so it is improper to be used to describe free electromagnetic field or photon field with wave nature. Free electromagnetic fields or photon fields should satisfy wave equation
\begin{equation}
\partial^2{A}_{\mu}=0
\end{equation}
In order to get it from Eq.(1), the Lorentz condition
\begin{equation}
\partial_{\mu}{A}_{\mu}=0
\end{equation}
is introduced. The spatial solution of Eq.(4) can be written as
\begin{equation}
A_{\mu}(x)=\int^{+\infty}_{-\infty}{a}_{\mu}(\vec{k})e^{ik\cdot{x}}{d}^3\vec{k}
\end{equation}
Put it into the Lorentz condition, we get
\begin{equation}
k_{\mu}{a}_{\mu}(\vec{k})=0
\end{equation}
This is a restriction condition about parameters $a_{\mu}(\vec{k})$. Only three $a_{\mu}(\vec{k})$ are independent among four.
\par
It is proved below that by introducing similar restriction condition, the motion equations of non-Ablian gauge fields and their solutions can also be simplified. The Lagrangian of non-Ablian gauge fields with zero masses is
\begin{equation}
L=-{1\over{4}}{F}^{\alpha}_{\mu\nu}{F}^{\alpha}_{\mu\nu}
\end{equation}
in which
\begin{equation}
F^{\alpha}_{\mu\nu}(x)=\partial_{\mu}{A}^{\alpha}_{\nu}(x)-\partial_{\nu}{A}^{\alpha}_{\mu}(x)+gf^{\alpha\beta\gamma}{A}^{\beta}_{\mu}(x){A}^{\gamma}_{\nu}(x)
\end{equation}
We can get the motion equations of free non-Abelian gauge fields from Eq.(8)
\begin{equation}
\partial_{\mu}{F}^{\alpha}_{\mu\nu}+gf^{\alpha\beta\gamma}{A}^{\beta}_{\mu}{F}^{\gamma}_{\mu\nu}=0
\end{equation}
Put Eq.(9) into (10), we obtain
\begin{equation}
\partial^2{A}^{\alpha}_{\nu}-\partial_{\nu}\partial_{\mu}{A}^{\alpha}_{\mu}+gf^{\alpha\beta\gamma}[(\partial_{\mu}{A}^{\beta}_{\mu}){A}^{\gamma}_{\nu}+A^{\beta}_{\mu}(\partial_{\mu}{A}^{\gamma}_{\nu})]
\end{equation}
$$+gf^{\alpha\beta\gamma}{A}^{\beta}_{\mu}(\partial_{\mu}{A}^{\gamma}_{\nu}-\partial_{\nu}{A}^{\gamma}_{\mu}+gf^{\gamma\rho\sigma}{A}^{\rho}_{\mu}{A}^{\sigma}_{\nu})=0$$
This is a complex non-linear equation, no solution is found up to now. Because it is not a wave equation, the non-Alebrian gauge fields seam to have no simple wave solution like Eq.(6).
\par
On the other hand, the middle Bosons described by non-Alebrian gauge fields are also micro-particles with wave nature. From this angle, the free non-Alebrian gauge fields should also have wave solutions. In fact, in the current perturbation calculations of practical problems, we have actually supposed that the wave functions of free non-Alebrian gauge particles have the simple forms like Eq.(6). For example, in the united weak-electric interaction theory, the interaction Hamilitonian between leptons and $W^{\pm}$ particles can be written as
\begin{equation}
{{ig}\over{2\sqrt{2}}}[W^{-}_{\mu}\bar{\psi}_{\nu}\gamma_{\mu}(1+\gamma_5)\psi_l+W^{+}_{\mu}\bar{\psi}_l\gamma_{\mu}(1+\gamma_5)\psi_{\nu}]
\end{equation}
By using the Feynman rule, the vertex function of interaction between leptons and $W^{\pm}$ particles in momentum space can be written as
\begin{equation}
{{ig}\over{2\sqrt{2}}}\gamma_{\mu}(1+\gamma_5)\delta^4(p_{l}+p_{\nu}-p_{w})
\end{equation}
In the formula, we actually acquiesce in that the wave functions of free $W^{\pm}$ particles have the simple wave solutions Eq.(6) similar to leptons. After the integral of transition matrix element is carried out in coordinate space, the formula (13) is obtained from Eq.(12). If the wave functions of free $W^{\pm}$ particles can not be written as the form of Eq.(6), we can not get vertex $\delta$ function, so that energy-momentum conservation in vertex processes can not be ensured and no any perturbation calculation can done.
\par
So the difficult position we face is that in order to let perturbation calculation becoming possible, we should suppose that the wave functions of free non-Alebrian gauge particles have simple form of the wave. On the other hand, according to the motion equation of free non-Alebrian gauge fields, the wave functions of non-Alebrian gauge particles have no simple form of the wave.
\par
This problem can be solved well by introducing a restriction condition in the motion equations of free non-Alebrian gauge fields, similar to introduce the Lorentz condition in electromagnetic theory. For this purpose, in Eq.(11) we let 
\begin{equation}
\partial_{\nu}\partial_{\mu}{A}^{\alpha}_{\mu}-gf^{\alpha\beta\gamma}[(\partial_{\mu}{A}^{\beta}_{\mu})A^{\gamma}_{\nu}+A^{\beta}_{\mu}(\partial_{\mu}{A}^{\gamma}_{\nu})]
\end{equation}
$$-gf^{\alpha\beta\gamma}{A}^{\beta}_{\mu}(\partial_{\mu}{A}^{\gamma}_{\nu}-\partial_{\nu}{A}^{\gamma}_{\mu}+gf^{\gamma\rho\sigma}{A}^{\rho}_{\mu}{A}^{\sigma}_{\nu})=0$$
The motion equation of free non-Alebrian gauge fields is simplified as
\begin{equation}
\partial^2{A}^{\alpha}_{\mu}=0
\end{equation}
In this way, the wave functions of free non-Alebrian gauge particles can be written as simple wave's forms
\begin{equation}
A^{\alpha}_{\mu}(x)=\int^{+\infty}_{-\infty}{a}^{\alpha}_{\mu}(\vec{k}){e}^{ik\cdot{x}}{d}^3\vec{k}
\end{equation}
Eq.(14) is just the restriction conditions of the motion equations of non-Alberian gauge fields. Putting Eq.(16) into (14), we get
\begin{equation}
\int{k}_{\mu}{k}_{\nu}{a}^{\alpha}_{\mu}(\vec{k})e^{ik\cdot{x}}{d}^3\vec{k}+igf^{\alpha\beta\gamma}\int\int(k_{\mu}+k'_{\mu}){a}^{\beta}_{\mu}(\vec{k}){a}^{\gamma}_{\nu}(\vec{k}')e^{i(k+k')\cdot{x}}{d}^3{k}{d}^3\vec{k}'
\end{equation}
$$+igf^{\alpha\beta\gamma}\int\int{a}^{\beta}_{\mu}(\vec{k})[k'_{\mu}{a}^{\gamma}_{\nu}(\vec{k}')-k'_{\nu}{a}^{\gamma}_{\mu}(\vec{k}')]e^{i(k+k')\cdot{x}}{d}^3{k}{d}^3\vec{k}'$$
$$+g^2{f}^{\alpha\beta\gamma}{f}^{\gamma\rho\sigma}\int\int\int{a}^{\beta}_{\mu}(vec{k}){a}^{\rho}_{\mu}(\vec{k}'){a}^{\sigma}_{\nu}(\vec{k}'')e^{i(k+k'+k'')\cdot{x}}{d}^3\vec{k}{d}^3\vec{k}'{d}^3\vec{k}''=0$$
Because $a^{\alpha}_{\mu}(k)$ has nothing to do with space-time coordinate $x$, we integrate the formula over $d^4{x}$ and we get
\begin{equation}
\int{k}_{\mu}{k}_{\nu}{a}^{\alpha}_{\mu}(\vec{k})\delta^4(k){d}^3\vec{k}+igf^{\alpha\beta\gamma}\int\int\delta^4(k+k')(k_{\mu}+k'_{\mu}){a}^{\beta}_{\mu}(\vec{k}){a}^{\gamma}_{\nu}(\vec{k}'){d}^3{\vec{k}}{d}^3\vec{k}'
\end{equation}
$$+igf^{\alpha\beta\gamma}\int\int\delta^4(k+k'){a}^{\beta}_{\mu}(\vec{k})[k'_{\mu}{a}^{\gamma}_{\nu}(\vec{k}')-k'_{\nu}{a}^{\gamma}_{\mu}(\vec{k}')]{d}^3\vec{k}{d}^3\vec{k}'$$
$$+g^2{f}^{\alpha\beta\gamma}{f}^{\gamma\rho\sigma}\int\int\int\delta^4(k+k'+k''){a}^{\beta}_{\mu}(\vec{k}){a}^{\rho}_{\mu}(\vec{k}'){a}^{\sigma}_{\nu}(\vec{k}''){d}^3\vec{k}{d}^3\vec{k}'{d}^3\vec{k}''=0$$
Take the integral of the first item about $d^3\vec{k}'$, the second and third items about $d^3\vec{k}'$ and the fourth item about $d^3\vec{k}''$. When $\vec{k}\rightarrow-\vec{k}$, $k_0$ is unchanged. Let $\bar{k}=(-\vec{k},ik_0)$, we get
\begin{equation}
-\delta_{\nu{0}}\delta(k_0){k}^2_{0}{a}^{\alpha}_{0}(\vec{k})-gf^{\alpha\beta\gamma}\int\delta(2k_0)2k_0{a}^{\beta}_0(\vec{k}){a}^{\gamma}_{\nu}(-\vec{k}){d}^3\vec{k}
\end{equation}
$$+igf^{\alpha\beta\gamma}\int\delta(2k_0){a}^{\beta}_{\mu}(\vec{k})[\bar{k}_{\mu}{a}^{\gamma}_{\nu}(-\vec{k})-\bar{k}_{\nu}{a}^{\gamma}_{\mu}(-\vec{k})]{d}^3\vec{k}$$
$$+g^2{f}^{\alpha\beta\gamma}{f}^{\gamma\rho\sigma}\int\int\delta(k_0+k'_0+k''_0){a}^{\beta}_{\mu}(\vec{k}){a}^{\rho}_{\mu}(\vec{k}'){a}^{\sigma}_{\nu}(-\vec{k}-\vec{k}'){d}^3\vec{k}{d}^3\vec{k}'=0$$
For the gauge particles with zero masses, $k_0=\mid\vec{k}\mid$, $k'_0=\mid\vec{k}'\mid$, $k''_{0}=\sqrt{\vec{k}^2+\vec{k'}^2}$. Then taking differential about $\vec{k}$ in the formula above, we get
\begin{equation}
\delta_{\nu{0}}{{\partial}\over{\partial\vec{k}}}[\delta(k_0)k^2_0{a}^{\alpha}_{0}(\vec{k})]+gf^{\alpha\beta\gamma}\delta(2k_0)\{2k_0{a}^{\beta}_{0}(\vec{k}){a}^{\gamma}_{\nu}(-\vec{k})-ia^{\beta}_{\mu}(\vec{k})[\bar{k}_{\mu}{a}^{\gamma}_{\nu}(-\vec{k})-\bar{k}_{\nu}{a}^{\gamma}_{\mu}(-\vec{k})]\}
\end{equation}
$$-g^2{f}^{\alpha\beta\gamma}{f}^{\gamma\rho\sigma}\int\delta(k_0+k'_0+k''_0){a}^{\beta}_{\mu}(\vec{k}){a}^{\rho}_{\mu}(\vec{k}'){a}^{\sigma}_{\nu}(-\vec{k}-\vec{k}'){d}^3\vec{k}'=0$$
The formula above is non-linear differential integral equation about parameters $a^{\alpha}_{\mu}(\vec{k})$ and is difficult to get its solutions. But in this way, the complexity of wave functions of non-Ablerian gauge particles is transformed into the complexity of parameters $a^{\alpha}_{\mu}(\vec{k})$ and the solutions of non-Abelrian gauge fields can be written as simple forms. There are $\alpha$ restriction equations, so only $3\alpha$ are independent among $4\alpha$ $a^{\alpha}_{\mu}(\vec{k})$. For the free non-Ablerian gauge fields with masses, the motion equation become
\begin{equation}
\partial^2{A}^{\alpha}_{\mu}-m^2_{\alpha}{A}^{\alpha}_{\mu}=0
\end{equation}
The form of restriction equations are unchanged, we only need to let $\mid\vec{k}\mid\rightarrow\sqrt{\vec{k}^2+m^2_{\alpha}}$ in Eq.(20). When there exist interactions, the relative interaction item can be added into the formula, but the form of restriction equation is still unchanged.
\par
Though the motion equations are simplified after the restriction conditions are introduced, the form of the Lagrangians of free non-Alberian gauge fields and interactions between gauge fields and other fields are unchanged. So after the solutions of motion equations of free non-Alberian gauge fields are written in the form of Eq.(16), the forms of interactions and practical calculation results are not effected. In fact, only in this way, we can use the Feynman rules in the calculation of perturbation theory, otherwise the calculations are completely impossible.
\end{document}